\documentclass[bibyear]{aa}
\usepackage[varg]{txfonts}
\usepackage{graphicx}
\usepackage{amssymb}

\usepackage{natbib,twoopt}
\usepackage[breaklinks=true]{hyperref}
\bibpunct{(}{)}{;}{a}{}{,}
  \newcommandtwoopt{\citeads}[3][][]{\href{http://adsabs.harvard.edu/abs/#3}%
    {\def\hyper@linkstart##1##2{}%
     \let\hyper@linkend\@empty\citealp[#1][#2]{#3}}}
  \newcommandtwoopt{\citepads}[3][][]{\href{http://adsabs.harvard.edu/abs/#3}
    {\def\hyper@linkstart##1##2{}%
     \let\hyper@linkend\@empty\citep[#1][#2]{#3}}}
  \newcommandtwoopt{\citetads}[3][][]{\href{http://adsabs.harvard.edu/abs/#3}%
    {\def\hyper@linkstart##1##2{}%
     \let\hyper@linkend\@empty\citet[#1][#2]{#3}}}
  \newcommandtwoopt{\citeyearads}[3][][]%
    {\href{http://adsabs.harvard.edu/abs/#3}
    {\def\hyper@linkstart##1##2{}%
     \let\hyper@linkend\@empty\citeyear[#1][#2]{#3}}}
\makeatother

\newcommand{\kms}{km\,s$^{-1}$}

\newcommand{\HII}{H{\sc ii}}
\newcommand{\sunn}{$_{\odot}$}

\newcounter{qub}

\begin{document}

\title {On variability of DDO68-V1, a unique extremely metal-poor LBV}

\author {S.A.~Pustilnik\inst{1}\thanks{E-mail: sap@sao.ru (SAP)}
	\and
	Y.A.~Perepelitsyna\inst{1}}

\institute{Special Astrophysical Observatory of RAS,\\ Nizhnij Arkhyz,
Karachai-Circassia 369167, Russia}

\date{Accepted 2025 January 29, Received 2024 November 29}

\abstract
{DDO68-V1 is a Luminous Blue Variable (LBV) star in the eXtremely Metal-Poor (XMP) galaxy DDO68.
It resides in the \HII\ region with 12+log(O/H) $\sim$ 7.1 dex, or $Z$ $\sim$ $Z$\sunn/40. Since DDO68-V1 is the only
known LBV with a so low initial metallicity, its in-deep study can give the hints for understanding the LBV
evolutionary stage and the nature of their powerful and highly variable mass loss in the very low-metallicity regime.}
{Our goal is to study the optical variability of DDO68-V1 during the last 36 years, with the emphasis on the
period of the last 8 years, after the LBV giant eruption. }
{We use our published results of monitoring in $B$, $V$, $R$ bands of the total flux of \HII\ region 'Knot~3',
containing the LBV,
along with photometry of the archive Hubble Space Telescope (HST) images, obtained in May 2010 and December 2017.
This data allow us to disentangle the variable light of DDO68-V1 and that of the underlying \HII\ region. }
{From all available photometry of Knot~3, we derive the $V$-band lightcurve of DDO68-V1 since 1988, with a higher
cadence during the years 2015--2023, when the lightcurve resembles that of S~Doradus.}
{The new data reveal the full range of DDO68-V1 absolute magnitudes $M_{\rm V}$ of [--5.9, --10.8]~mag.
The LBV variations after the fading of the 'giant eruption' show the unusually large amplitude of
$\delta~V$ $\gtrsim$ 3.0--3.5~mag  on the time-scale of $\sim$1--1.5 year. The apparent changes
of the integrated $B-V$ colour of Knot~3 are consistent with the expected colour variations of the LBV in course of
the S~Doradus 'normal eruptions'.
These data, along with spectra of DDO68-V1, demonstrate the need for a higher-cadence photometry
of DDO68-V1, in order to probe the possible periodicity in its lightcurve and binarity of the object.
}


\maketitle
\nolinenumbers

\section[]{Introduction}
\label{sec:intro}

Luminous blue variable (LBV) stars belong to the quite rare class of the evolved massive stars.
The traditional view on LBVs considers them as a relatively short (several 10$^5$ or less years)
transient and rather unstable stage of massive star evolution from the main
sequence hydrogen burning O-stars to the core-helium burning Wolf-Raye (WR) stars \citep{HD94}.
They display evidence of the strong mass loss via powerful non-stationary winds and eruptions
that result in formation of circumstellar shells with radii of a fraction of pc and more
\citep[e.g.][and references therein]{Weis2020,Kniazev2016}.
More resent data indicate that LBVs can evolve as the direct precursors of supernova
\citep[e.g.][and references therein]{Petrov2016}. Moreover, as \citet{Smith2017} argues,
LBVs originate in massive binaries and are gainers of the binary co-evolution. The recent results of
\citet{Mahy2022} give more support to this concept.

The great majority of known and candidate LBVs belong to the Milky Way, M31, M33,
Magellanic Clouds and other rather massive galaxies, and, hence, have the solar
or several times lower metallicities. Meanwhile, the study of the most metal-poor massive stars in
the local Universe attracts much attention as related to our understanding the processes of galaxy evolution and star
formation feedback in the early Universe.
This direction is currently supported by several new projects
\citep[e.g.][]{Garcia21, Gull22, Lorenzo22, Vink_ULLYSSES}.

Probing the evolution of such massive stars via the  properties of the most metal-poor LBVs is especially
suitable thanks to the two factors. First, due to their highly non-stationary mass loss, their properties
should be more sensitive to the model assumptions, and thus, to be helpful in the comparison of model predictions
with the real very-low-metallicity massive stars.
And second, their observed relatively short time-scales (years-decades) for the substantial changes of
their evolutionary status, including giant eruptions with the large mass loss, the impostor and SN explosions
\citep[e.g.][]{Petit06, Smith2016, Guseva2022, Aghak2023a}, make them especially valuable observational targets
for monitoring programs.

The issue of the extremely metal-poor massive star evolution remains a crucial for understanding of
galaxy formation and evolution for the time  of $\lesssim$1~Gyr since the Big Bang
\citep[e.g.][and references therein]{Eldridge22}. While the state-of-art stellar evolution models, including those
with the fast rotation, have advanced during the last years
\citep[e.g.][and references therein]{Szecsi15, PARSEC, Sanyal2017, Vink2022}, there is no the direct comparison
of the model predictions with properties of real extremely metal-poor massive stars reported. The main reason
is the lack of such stars in the local Universe which would be accessible for sufficiently detailed studies.
The qualitative progress in the accessibility of the faint targets is expected with
the coming next generation of the extremely large optical telescopes.

As a preparatory step for the future detailed studies, the search for such rare massive stars and examination of
their available properties, such as variability, appear timely and necessary.

The Lynx-Cancer void \citep{PaperI} galaxy DDO68 (UGC5340) is known as a peculiar
morphology dIrr \citep[e.g.][]{DDO68, DDO68LBV, Ekta08, Annibali2019b} with a number
of prominent young star-forming (SF) regions.
Several of them have the almost record-low metallicity: 12+log(O/H)=6.98--7.10~dex.
Most of its SF regions are found at the periphery, mainly in the 'Northern ring' and the 'Southern tail'
\citep{DDO68, IT07, Annibali2019b}. In the repeat
observations of DDO68, the unique Luminous Blue Variable star (LBV) was
discovered in one of the most metal-poor \HII\ regions in the local Universe
\citep{LBV,IT09}.

In this Letter, we present and discuss the lightcurve of the LBV DDO68-V1 (the LBV, originated
from the gas with metallicity of $\sim$$Z$\sunn/40), which
clearly evidence to its recent 'giant eruption' and the light variations afterwards, resembling
the S~Doradus type normal eruptions.

The lay-out of the Letter is as follows. In Sect.~\ref{sec:obs}, we present
all the used ground-based observational data on the aperture photometry
as well as the direct photometry of DDO68-V1 at the Hubble Space Telescope (HST) images in May~2010 and December~2017.
In Sect.~\ref{sec:results}, the main results of data processing and analysis
are presented. Sect.~\ref{sec:dis} is devoted to discussion of new results,
their comparison with the previous data and understanding them in a wider
context. In Sect.~\ref{sec:summ}, we summarise the new results and draw
the main conclusions. The details of the data and their analysis and
Tables with the original photometric series of Knot~3 and the derived magnitudes of
the LBV are presented in the 'Supplementary materials' as Appendices A, B and C.
The distance to DDO68 is adopted according to the 'Tip or Red Giant Branch' (TRGB) based estimate from
\citet{Makarov17} of 12.75~kpc. 
The latter is close to that from \citet{Sacchi16, Annibali2019a}.
The respective scale is 62~pc in 1\arcsec.

\section[]{Observational data}
\label{sec:obs}

\subsection{Ground-based data}
\label{ssec:ground}

The ground-based observations and data processing as well as the results of photometry
of region Knot~3 (among the other five regions) are described in detail in
\citet{DDO68NR}. They are also briefly summarised in 'Supplementary materials' (Appendix~B).
Our photometry was obtained in 35 nights during the period of March 2016 to October 2023,
mostly with SAO RAS 6m (BTA) and 1m telescopes and with 2.5m telescope of Caucasian Mountain Stations
of Moscow State University.
It was complemented by a few our earlier data from BTA and by the data collected from the archive images
of ten telescopes worldwide. The earlier and archive data were reanalysed in the same way as our new
observations. Therefore, the results in Table~\ref{tab:photo_Kn3} are slightly different from those in
Table~2 of \citet{DDO68LBV}. More details are given at the end of Appendix~B.

The aperture photometry of DDO68 \HII\ region Knot~3 on all images
 was based on the local standard stars.
Half-dozen sufficiently bright stars ($g = 17.9^m - 20.5^m$) were selected
in the SDSS \citep{DR7} images in the vicinity of DDO68-V1.
Their $g,r,i$ magnitudes were transformed to the Johnson-Cousins $B,V,R$
magnitudes according to the \citet{Lupton05} relations.

\subsection[]{HST data}
\label{ssec:HST}

The HST data used for this work, are the deep images in filters F606W and F814W, obtained in May 2010
and December 2017, on the programs ID 11578 (PI: A.Aloisi) and ID 14716 (PI: F.Annibali), respectively.
We use the photometry and the derived $V$, $I$ magnitudes of the fifty most luminous stars in the 2010 HST
Advanced Camera for Surveys (ASC)
image as presented in \citet{DDO68LBV}. The images in December 2017 were obtained with the Wide Field Camera 3,
twice in each of the filters. The total exposure times for both filters were of 7280~s for one pair, and 5116~s
-- for the other pair. The lag between the images in F606W was 13 days, and for images in F814W -- 19 days.
The photometry of objects of interest in the 2017 HST image was performed in this work,
with all details being presented in 'Supplementary material' (Appendix A). Briefly, we perform the two main procedures.

The first one is the two-Gauss fitting of the 1d scan through the image of the close 'pair' at the position of DDO68-V1.
This allows us to determine the relative fluxes of the two stars in the 'pair' and to connect their position to the
nearby stars. This allows us, after comparison of the LBV position in the HST image in May 2010, to fix that in
December 2017 the LBV was the fainter of these two stars.

The second procedure is to determine the $V$ and $I$ apparent magnitudes of the above stars. This is done via the aperture
photometry of the 'pair' and of 35 of 50 supergiants from \citet{DDO68LBV}, selected as clearly isolated of other stars,
so that the aperture photometry should be reliable. These 35 stars with known $V$, $I$
magnitudes are used as the local standards in the 2017 HST images in order to determine the respective zero-points in
the both $V$, $I$ bands. Applying these zero-points to the instrumental magnitudes of the total light of the 'pair', we
obtain its total $V$, $I$ magnitudes.
Having the flux ratio from the above two-Gauss fitting, we finally derive the $V$ magnitudes of both stars and their colour
$(V-I)$. We also tried to use the DOLPHOT PSF photometry, but due to the systematical differences between the two
epochs, we choose the current variant. See details in Appendix~\ref{sec:twogauss}.

\section[]{RESULTS}
\label{sec:results}

The main results of this study are presented in more detail in the 'Supplementary materials'. Here we summarise
them briefly.

\subsection[]{Results from HST images}
\label{ssec:HSTresults}

As told in the previous section, we use the HST images obtained in December 2017. We decompose the non-stellar
images of the object at the position of DDO68-V1 with the two-Gauss fitting and get the relative fluxes of the
two adjacent stars.
Then we perform the aperture photometry of this non-stellar object. Finally we transform its instrumental magnitudes
to the apparent $V$ and $I$ ones, using the zero-points in each of the HST images as explained in the previous section.
See Appendix~\ref{sec:twogauss} for more detail.

The brighter star of the two, with $V$ = 23.99$\pm$0.05, M$_{\rm V,0}$ = --6.90~mag and the colour $(V-I)_{\mathrm 0}$ =
--0.109$\pm$0.115 can be assigned to B5-B8 supergiant \citep[e.g.][]{Wegner94, Ducati01}.
For DDO68-V1 at this epoch, we derive $V$ = 25.00$\pm$0.12~mag.
With the absolute value of M$_{\rm V,0}$ = --5.89~mag and the colour $(V-I)_{\mathrm 0}$ = --0.31$\pm$0.34,
the LBV near the minimum is similar to an early (about B1) B supergiant \citep{Ducati01}. However, due
to rather large uncertainty of its colour index, the LBV effective temperature in this phase can range between
that for O5 to late B supergiant \citep{Wegner94,Ducati01}. The better accuracy of the
LBV colours is necessary to constrain its spectral class near the minimum.
 The other direct estimate of the LBV magnitude ($V$ = 20.05) in the May 2010 HST image \citep{DDO68LBV} implies
 the minimal registrated amplitude of the LBV light variations of $\delta~V$$\gtrsim$5.0~mag.

Since we have for the December 2017 HST data two independent images for each of the filters, we can check possible
light variations of DDO68-V1 on the time scales of 2--3 weeks. The respective upper limit for variations of DDO68-V1
during this period is $\delta~V \lesssim $0.03~mag. See  more details in Appendix~\ref{sec:twogauss}.

\subsection[]{DDO68-V1 lightcurve from the ground-based photometry}
\label{ssec:lightcurve}

In Table~\ref{tab:photo_Kn3}, we summarise all our photometric data on the
integrated magnitudes of Knot~3.
In Col.~1, the date of observation is given in the format YYYYMMDD. In Cols.~2 and 3,
we give the total $B$-mag and its adopted uncertainty $\sigma_{\rm B}$.
In Cols.~4 and 5, we give the total $V$-mag and its $\sigma_{\rm V}$. Cols.~6 and 7
give the $R$-magnitudes and their errors.
In Col.~8, we give the information on the used telescope.

We suggest that, at first approximation, the total light of Knot~3 is composed of the (almost) non-variable light
of the related \HII\ region and of many unresolved stars within the used aperture (radius of 2.5 arcsec), and the
variable light of DDO68-V1. In this approximation, we can get the flux of the LBV at the respective epochs via the
subtraction
of the non-variable light of the 'underlying complex object' from the measured total light of Knot~3.
The magnitude of this non-variable object can be determined iteratively, based on the faintest measured $V$-mag of
Knot~3 in Table~C.1.

In particular, on the epoch 2005 Jan 12, when the first spectra of Knot~3 were obtained, they demonstrated the
purely \HII\ region emission lines, with no indication on the LBV emission contribution \citep{DDO68}.
Its $V$-mag of 20.19$\pm$0.013 in this date defines the upper limit to the light of the non-variable object. Its real
level should be at least several 0.01~mag fainter, in order the residual flux of the LBV remained at the reasonable
level. That is, its M$_{\rm V,0}$ is brighter than --5.8~mag, corresponding (with the account for A$_{\rm V}$ =0.36
within the \HII-region) to $V$(LBV) $\sim$25.0). The iteration procedure leads to the level of this
non-variable 'background' light at $V_{\rm back}$ = 20.23 -- 20.24~mag. See more details in Appendix~\ref{sec:subtract}.
In Fig.~\ref{fig:LBV_eruption}, we show the resulting $V$-band lightcurve of DDO68-V1 in the three time intervals.

The effect of the choice of V$_{\rm back}$ is illustrated by green (V$_{\rm back}$=20.23) and red (V$_{\rm back}$=20.24)
symbols.
The direct estimates of $V$(LBV) in the HST images for epochs of May 2010 and December 2017 are overplotted as the
black dots.
As one can see, a bit fainter level of the background component affects substantially the minima of the LBV lightcurve,
reducing somewhat the amplitude of variations after the giant eruption. However, if we try to further reduce the level
of V$_{\rm back}$, say to $\sim$20.25~mag, all the ground-based minima of the LBV lightcurve shift upwards by about
a half-magnitude or more. This leads to the situation, when the HST-defined minimum becomes too deep in comparison to
the other minima.
While we can not exclude such a case completely, it is much more likely that the HST-based minimum
is similar to the minima, detected from the ground-based photometry. Therefore, we consider the level of
V$_{\rm back}$=20.23--20.24~mag for the non-variable background as close to the real.

In Table~\ref{tab:photo_V}, we present the derived $V$-band values for sum (LBV+BSG) (Col. 2) with their errors
(Col. 3), which are transferred from the errors of $V$(Knot~3) in Table~C.1.  In Cols. 4 and 5, the estimates
 $V$-mag and their errors are given for the LBV itself, after subtraction the light of the neighbour BSG.
The latter errors are transferred from the errors of $V$(LBV+BSG).  In Fig.~\ref{fig:LBV_eruption}, the large
errors of $V$(LBV) ($\sigma_{V, LBV} > $ 1mag) near the minima of the lightcurve are artificially reduced to keep the
frame range more compact.

\subsection[]{DDO68-V1 colours}
\label{ssec:colors}

For a part of the ground-based observations, we have the simultaneous measurements in two or
three bands ($B, V, R$). So, we can probe the possible trends of the colour variations versus the total
luminosity of Knot~3, which, in turn, are correlated with variations of DDO68-V1.
See  Appendix~\ref{sec:color.var} and Fig.~\ref{fig:K3_colours}.

In brief, there are indeed some systematic trends in the colour variations of Knot~3 in the period after
the giant eruption. The colours of Knot~3 change substantially from the minima
to the local maxima of the lightcurve.
The reddening of $(B-V)$ colour with the increase of Knot~3 brightness can be consistent with
the commonly adopted cooling of the LBV 'photosphere' from a blue supergiant to F supergiant.
However, $(V-R)$ colour of Knot~3 gets bluer near the local maxima. This can be related
to variable contribution of H$\alpha$ emission in the LBV. See Appendix C for more detail.

\section[]{DISCUSSION}
\label{sec:dis}

As said above, DDO68-V1 is the only known LBV with the initial extremely-low metallicity
($Z$(gas)$\sim$$Z$\sunn/40). Therefore, the understanding its properties is very important for comparison with
model predictions of the very low-$Z$ massive star evolution. However, its brightness, excluding the period of the
giant eruption event, is mostly too low for its high-resolution spectroscopy. So, the detailed spectral study of this
LBV will take special efforts and the use of either HST, or the upcoming extremely large telescopes.

The situation, however, is much better for studying its light and SED variations. The LBV is caught during
the giant eruption, and also is being registrated afterwards, during the period of about eight years,
when it displayed the light variations resembling those of S Doradus type variables.
We present the lightcurve of DDO68-V1 in $V$-band in Fig.~\ref{fig:LBV_eruption} and briefly discuss its features below.
We mention also the $B$-band photometry of \citet{Bomans11} for Knot~3 [12 epochs in 1955--2007],
which, however, shows the $B$ magnitudes substantially fainter relative to ours in the close epochs.

The main features of the presented lightcurve are: 1) the 'giant eruption', which reached the brightness peak of
$V \sim$ 20.0~mag
(corresponding to $M_{\rm V} \sim$ --10.8~mag) near the middle of 2010, and 2) the quasi-regular light variations after
the giant eruption, in the period since 2015. In course of the giant eruption, the brightness of DDO68-V1 increased
by $\sim$2.2~mag relative to the median level ($V\sim$22.2~mag), and by $\sim$5~mag
relative to the deep minima of the lightcurve. The duration of the giant eruption is about 6.5 years, if we define it by the
lightcurve intersections of the median line near 2006.5 and 2013. The first signs of the broad emission lines, related to
the giant eruption, are detected only in January 2008.

It is interesting that both, the beginning and the 'end' of the giant eruption, are rather close on time to two deep
minima in the light curve, in the beginning of 2005 and in 2015.
Another curious feature of the LBV spectral behaviour in course of the giant eruption fading, was the detection in its
spectrum of the substantially diminished broad H$\alpha$ in April 2016 by \citet{Guseva2022}. This appeared in a year
after the deep minimum in 2015, and in fact, very close to the local maximum of the subsequent large variations.
While this broad line could be some residual emission of the strongly faded shell, this and alternative options
take a more careful analysis. This broad component of H$\alpha$ faded completely, however, in two more years (April 2018).

The light variations since 2015 are not perfectly sampled. However, the main their features can be uncovered
and summarised as follows. Of 32 DDO68-V1 magnitude estimates for this time range, only five points relate to the evident
minima, with the magnitude range from their nearest maximum of $\sim$3.0--3.5 mag. Probably two more minima are present in
the lightcurve with the amplitudes of $\sim$1~mag. The characteristic times of the fall to a minimum and the rise after it
range from $\sim$0.2 to $\sim$1~year. While one can not exclude the effect of small statistics, nevertheless,
all five minima during this period are registrated by only a singular measurement, what indicates their relatively
short duration, of $\lesssim$0.5~year.
A good example is the deep HST-based minimum in December 2017. This combines 4 HST measurements in $V$ and $I$ filters
during 3 weeks and the adjacent intermediate brightness point in mid-November 2017. The two adjacent local maxima,
at the end of May 2017 and in Feb. 2018, are separated by only 9 months.

In difference to this, the local maxima can be defined by several points with the total duration of about 1~year (as in
periods of March 2016 - June 2017, Feb. 2018 - Feb. 2019, and Dec. 2022 - Oct. 2023). Clearly, the higher-cadence data
with the more regular time coverage would be helpful to better characterise the lightcurve of DDO68-V1.
In general, from the presented picture of the whole lightcurve, during the last 36 years since 1988, the LBV is
in the 'agitated' state. Indeed, of 44 shown LBV magnitudes, only 6 are close to the minima of its lightcurve,
corresponding to the quiescent state.
Despite the amount of photometric and spectroscopic data for DDO68-V1 is modest, it is useful to make a
preliminary comparison with other well studied LBVs with the giant eruptions or with well visible S~Dor variability.

There are two other LBVs in galaxies with the nearest (but still much larger) metallicity to that of DDO68.
These are LBVs in PHL293B and NGC2363-V1, with 12+log(O/H) = 7.71 and 7.90~dex.
Both these LBVs were caught during their giant eruptions.

For LBV in PHL~293B, with the maximum $M_{\rm V} \sim$ --12~mag, the giant eruption was lasting at least
$\sim$16 years \citep{Guseva2022}.
The broad component of H$\alpha$ was detected at first time in the SDSS spectrum in 2001.
In 2019, based on the disappearance of the broad H$\alpha$, this giant eruption is claimed to be faded
\citep{Allan2020}.  However, according to the spectrum, obtained by \citet{Guseva2022} in November 2020,
the broad H$\alpha$ is still present.
No information is published on the optical variability of this LBV before the giant eruption.

NGC2363-V1 was observed with the HST in both, spectral and photometric modes, during the period of its
maximal light, when it brightened by $\sim$3.5--4~mag and reached $M_{\rm V} \sim$--10.5~mag
\citep{Drissen01, Petit06}. The LBV brightened from the 'low state' to the 'flat' part of the giant eruption
in about 2 years, and then experienced small fluctuations at this level with the very slow fading in visual, but
brightening in UV, during $\sim$10 years (1995--2005).
Since no information is published on its lightcurve after 2005, neither the duration of the giant eruption
is known, nor the type of NGC2363-V1 lightcurve after its end.

Thus, the duration of the giant eruption in DDO68-V1 is two-three times shorter then for the two 'similar' LBVs.
As for the type of variability after the giant eruptions, no such information is available.

The lightcurve of DDO68-V1 after the giant eruption, in years 2015-2023, is better sampled and can be,
in principle, compared with the lightcurves of the known LBV stars, monitored during decades.
One of such examples is a well-known S-Doradus variable  AG~Car.
Its  about 30 years-long lightcurve was presented by \citet{Sterken2003}. AG~Car displays the
'short-term' ($\sim$1~year) variations with amplitudes of $\sim$0.5--1.5~mag  and the underlying
long-wave variations with the characteristic time of $\sim$15 years. The analysis of its variability reveals
period of P$\sim$370~days.

If we compare the 'regular' light variations in DDO68-V1 on the 1-year time-scale
with the lightcurves of the other known LBVs of S-Doradus type
\citep[e.g. AG~Car, HR~Car, R127, GR290][]{Sterken2003, Smith2020, Walborn2008, Polcaro2016},
we find that its amplitude of 3.0--3.5 mag much exceeds the typical amplitudes of the known LBVs, of $\sim$1--2 mags.
There are two factors, which could be related to the observed elevated amplitude of DDO68-V1 variations.
The first one can be due to the post-giant-eruption processes in the LBV, so that the amplitude will decrease in
years - decades.
The second factor can be related to the very low metallicity of the LBV stellar material.
It is also possible that both factors contribute to the enhanced amplitude of DDO68-V1 S~Dor type variations.

In addition, the 'regular' variations of DDO68-V1 after the giant eruption, on the time-scale of
$\sim$1 year, may resemble the unusual high-amplitude
and a rather short duty cycle ($\sim$200~days) behaviour of LBV in NGC3432 (SN impostor SN~2000ch)
\citep{Pastorello2010, Aghak2023a}. For this rare type of variability, \citet{Aghak2023a} find very clear evidence
of periodicity, which indicates the effect of periastrons in a binary system.

\section{Summary and conclusions}
\label{sec:summ}

Summarising the presented results and the discussion above, we emphasize the feasibility
of the ground-base photometry of DDO68-V1. While the conducted observations were of rather low cadence and not of the
best quality as they could be in the case of a more dedicated program, they resulted in the interesting outcome
and paved the way to a more advanced study of this unique LBV star. To check the possible binarity, that is
the hidden periodical component in the LBV light variations, one needs in a two-week or smaller cadence of observations.
Its monitoring in several broad-band filters with the typical accuracy
of $\lesssim$0.01~mag, should be adequate for getting a substantially improved observational material and the deeper analysis
of the LBV properties.
The correlated variations in several filters should ease the search for periodicity and allow one to study the
colour variations.
As our work demonstrates, this task can be accomplished with telescopes of 1--2 metre class.
Besides, since in the local maxima of the 'S Doradus-type' lightcurve, the LBV reaches the brightness level of
$V \lesssim$ 22~mag, one can use big telescopes to perform the spectroscopy of the LBV in these phases.

Our conclusions can be formulated as follows:

\begin{enumerate}
\item
The unique LBV DDO68-V1 was discovered in January 2008 in
the void galaxy DDO68, in the \HII-region with the almost record-low
metallicity of 12+log(O/H)$\sim$7.1. We update its earlier published
scarce lightcurve for the period of 2005 - 2015 from \citet{DDO68LBV},
adding our fresh (since the year 2016) SAO 6m and 1m telescopes and CMO 2.5m
telescope photometry of Knot~3 and the photometry from archive images
with ten different telescopes over the period of 1988 - 2013. The present
photometry over the last 36 years during 44 epochs reveals that LBV is
rather active during this period, leaving in the 'low' state only during six 'short'
intervals with the duration of $\lesssim$ 1~year.
\item
All accumulated data allow us to determine the reliable amplitude of this LBV lightcurve.
In $V$-band, it varied during the last three decades in the range of $\sim$20.0 to
25.0~mag or fainter. This corresponds to $M_{\rm V}$ range of $\gtrsim$ --5.9 to $\sim$--10.8~mag.
The major brightening of the LBV took place in 2009--2011, with the subsequent fading by $\sim$2~mag
by 2013. Combining this brightening with the uprise of the broad emission lines and the P~Cyg-type
absorptions with terminal velocity of $\sim$800~\kms, this event is classified as
the giant eruption,  which is related to the strong episodic mass loss, similar to the $\eta$~Car phenomenon.
\item
The monitoring of DDO68-V1 during 8 years after the end of the giant eruption reveals
'regular' light variations,  resembling those of S~Doradus, considered as typical of LBVs.
Thus, this data hints that DDO68-V1 indeed demonstrates the signs of the bona fide LBVs.
However, the amplitude of these 'regular' variations reaches of $\sim$3--3.5~mag,
substantially larger than for the known LBVs with the well sampled lightcurves. 
\item
The results of the study of DDO68-V1, presented in this work, are very promising, but still,
the statistics of variability is rather poor. Taking into account the unique status of this XMP LBV,
in order to get the deeper insights in its properties, we suggest a wide collaboration of its high-cadence
(1.5--2 weeks) multiband monitoring during the next several years. This will allow one to better understand the
characteristic times and amplitudes, to probe its potential binarity via the uncovered periodic signal
(probably hidden due to the internal LBV processes), and to clear up the trends in its colour changes.
This data will pave the way to the better understanding of the massive evolved stars with the lowest metallicities,
the analogs of massive stars in the early Universe.
\end{enumerate}

\begin{acknowledgements}

The results of this research are obtained with the financial support of grant No. 075-15-2022-262
(13.MNPMU.21.0003) of the Ministry of Science and  Higher Education of Russian Federation.
Observations at the 6-m telescope BTA are supported by funding from the
Ministry of Science and  Higher Education of Russian Federation
(agreement No~14.619.21.0004, project identification RFMEFI61914X0004).
We thank A.~Vinokurov and A.~Kniazev for their useful comments on the earlier version
of the paper. We also thank the anonymous referee for their comments.
We acknowledge the use of the SDSS database.
Funding for the Sloan Digital Sky Survey (SDSS) has been provided by the
Alfred P. Sloan Foundation, the Participating Institutions, the National
Aeronautics and Space Administration, the National Science Foundation,
the U.S. Department of Energy, the Japanese Monbukagakusho, and the Max
Planck Society. The SDSS Web site is http://www.sdss.org/.
The SDSS is managed by the Astrophysical Research Consortium (ARC) for the
Participating Institutions.
\end{acknowledgements}


\begin{appendix}



\section{HST images of DDO68-V1 in 2010 and 2017. Deblending light of the neighbour supergiant.}
\label{sec:twogauss}

\begin{figure}
  \centering
 \includegraphics[angle=0,width=7.0cm, clip=]{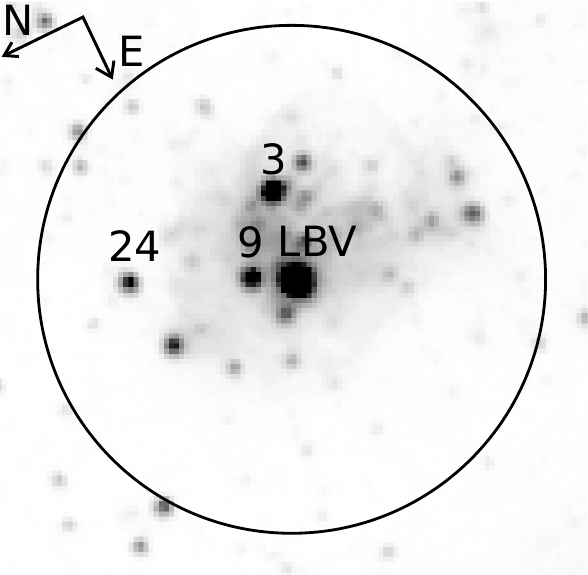}
 \includegraphics[angle=0,width=7.0cm, clip=]{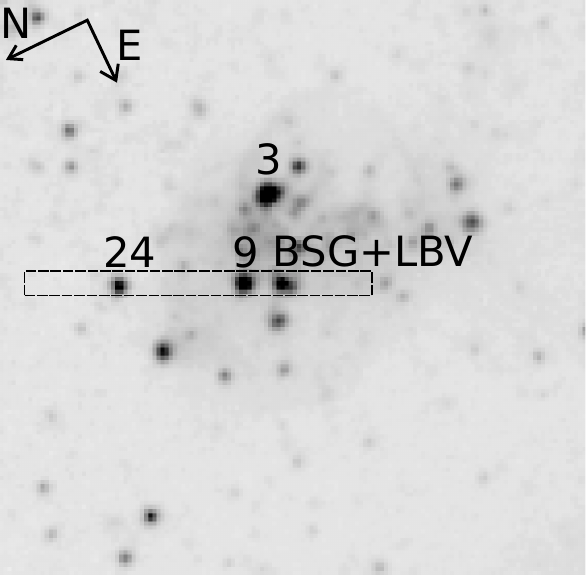}
  \caption{
   Part of the HST image of DDO68 in F606W filter, centred on the SF region Knot~3 with the LBV DDO68-V1 near its centre.
   The aperture, used for the ground-based photometry, is superimposed as a cycle with $D_{\rm aper} =$ 5\arcsec.
   Arrows show directions to the North and East.
{\bf Top panel:}   Epoch of May 2010 (Program ID 11578, PI: A.Aloisi).
{\bf Bottom panel:}  Epoch of December 2017 (Program ID 14716, PI: F.Annibali). The shown rectangular with width of 7 pixels
  (0.28 arcsec) is extracted and its light is summed-up along Y direction to perform 1D two-Gauss fitting of DDO68-V1
   region. See details in Appendix~\ref{sec:twogauss}.}
	\label{fig:knot_3_V}
 \end{figure}

In both filters, F606W and F814W, there is a clear indication on the asymmetric appearance of the object at
the position of DDO68-V1 in the HST images on the epoch of December 2017. The position of DDO68-V1 is defined at
the HST image on the epoch of May 2010 (Fig.~\ref{fig:knot_3_V}, top panel), when the LBV was near the peak of its brightness.
We rotate the original HST image by $\sim$120\degr\ counterclockwise, so that the direction from the LBV
to stars No.~9 and No.24 would coincide with X axis. We then extract the rectangular box along the X direction, as shown
in Fig.~\ref{fig:knot_3_V} (bottom panel) (the width of 7 pixels)
and summ-up the signal over these 7 pixels, to produce an 1D scan of the region of interest. Its width corresponds
to $\sim$2.5~FWHM of a stellar image.

We then perform the procedure 'blendfit' from \mbox{MIDAS}, to fit the 1D brightness profile of the object near the LBV
position by two Gaussians. We fix the width of the Gaussian to be equal to the width of two other stars visible on
the 1D scan (No.~9 with $V$=23.38~mag and No.~24 with $V$=24.30~mag, where their nomenclature is from \citet{DDO68LBV}),
namely at 2.85 pixels (or $\sim$0.11 arcsec).
We than vary the relative amplitudes and the distance between the two components to get the best $\chi^2$ parameter
and the least value of residuals (see Fig.~\ref{fig:blendfit_LBV}).

\begin{figure}
  \centering
\includegraphics[angle=-90,width=8.0cm,clip=]{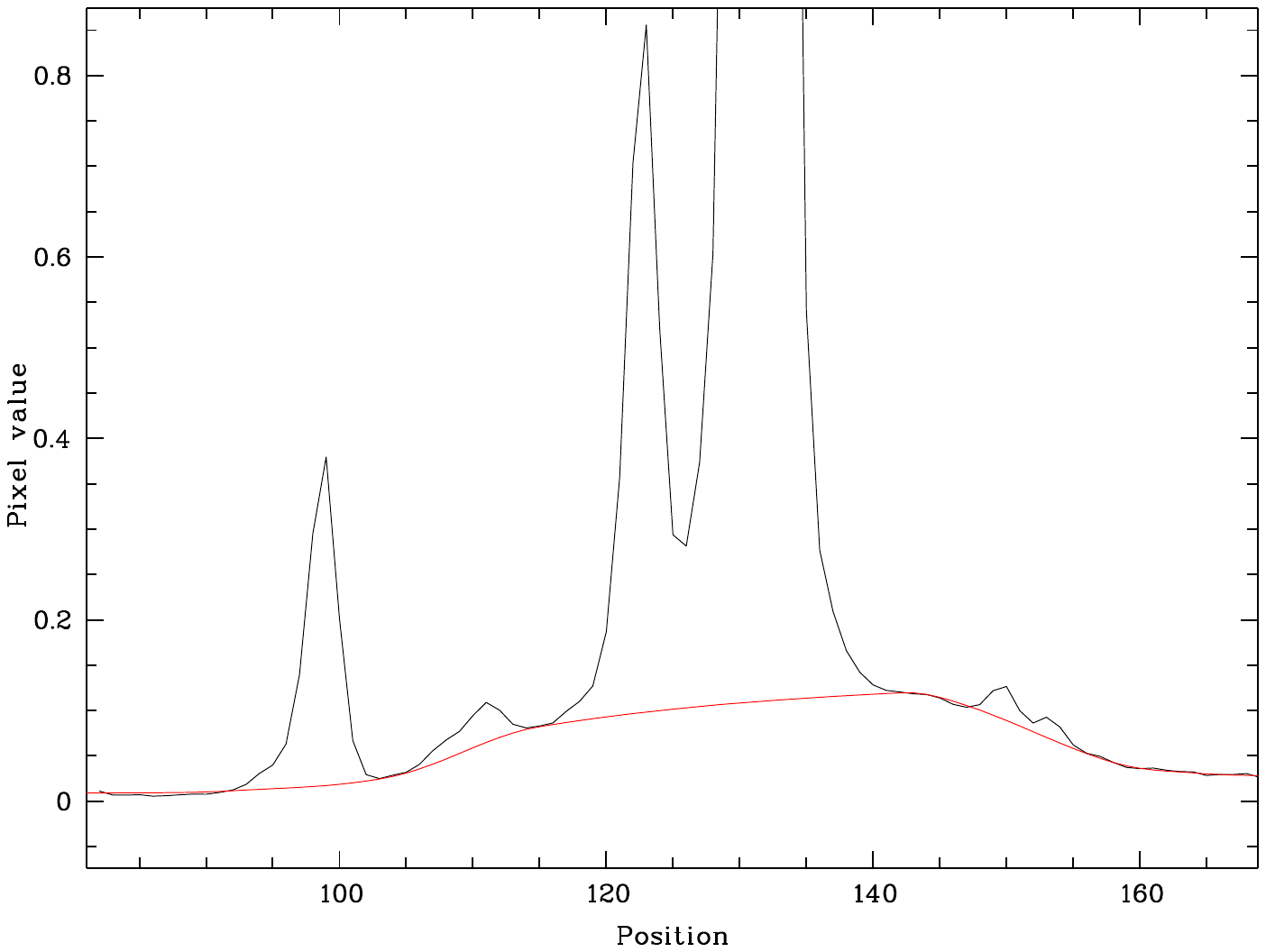}
\includegraphics[angle=-90,width=8.0cm,clip=]{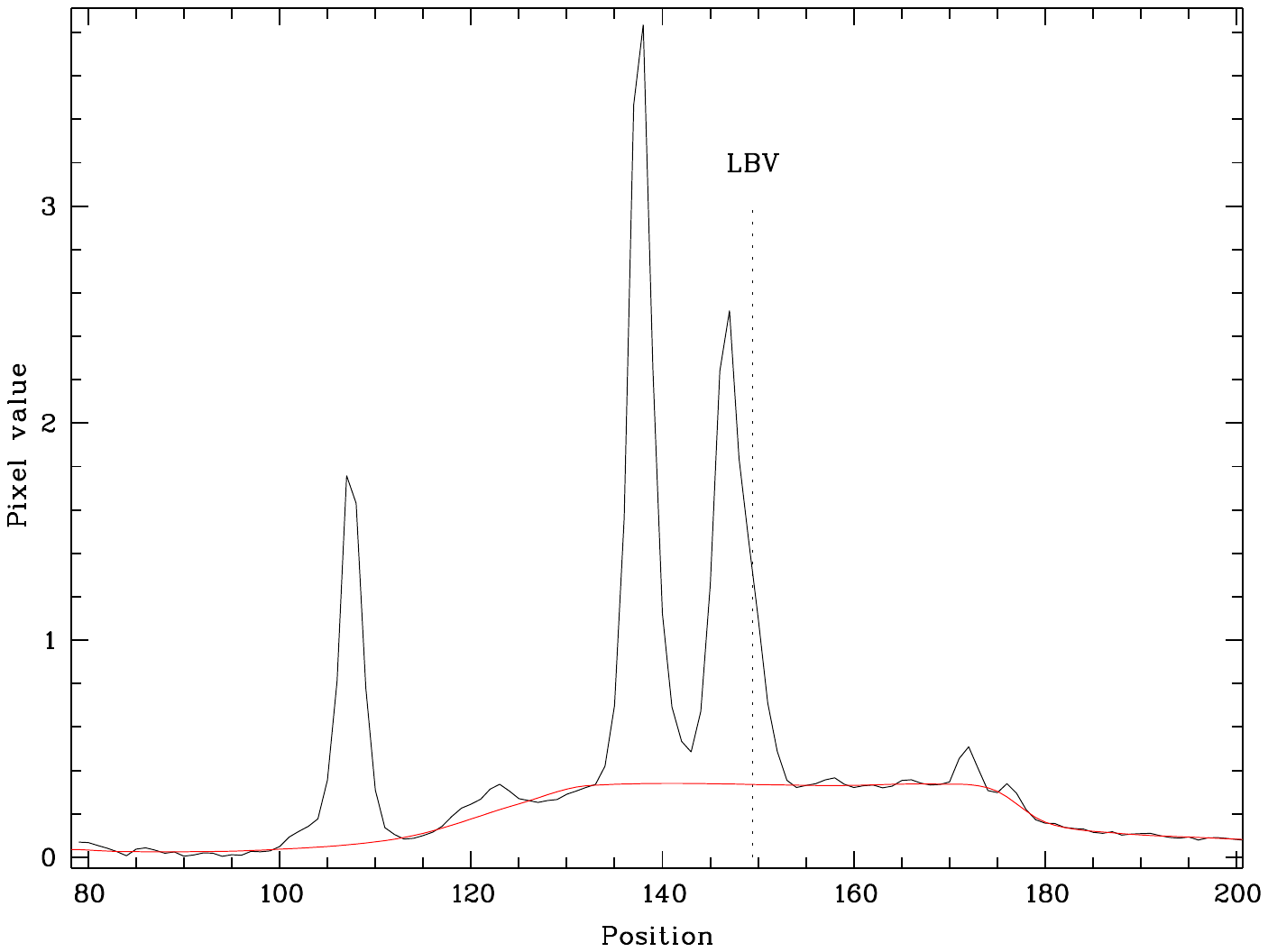}
  \caption{{\bf Top panel:} 1D scan of brightness in V-band near DDO68-V1 on the epoch of May 2010.
{\bf Bottom panel:} Same as in the left-hand panel, but for the epoch of December 2017. Red solid line shows the level of
   the underlying background, above which the two-Gauss  fitting was performed, as shown in
   the left panel of Fig.~\ref{fig:blendfit_LBV}.
}
	\label{fig:HST_2_epoch}
 \end{figure}

\begin{figure}
  \centering
 \includegraphics[angle=-0,width=3.5cm, clip=]{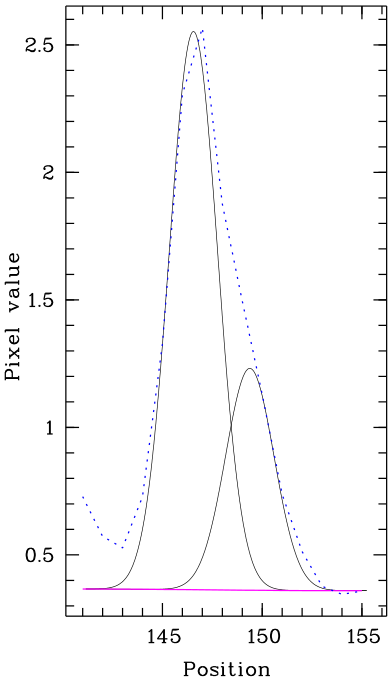}
 \includegraphics[angle=-0,width=3.5cm, clip=]{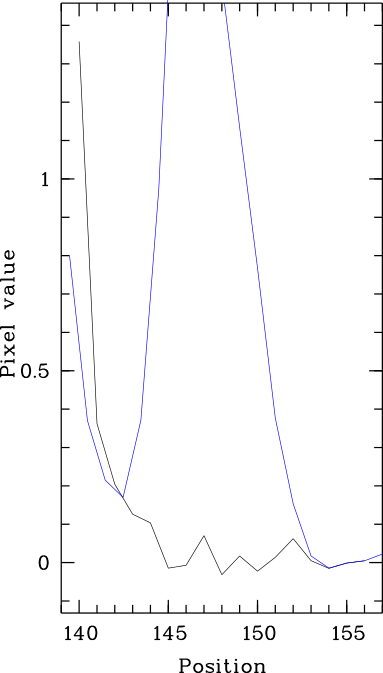}
  \caption{
{\bf Left-hand panel:} Two-Gauss fitting to the $V$-band brightness profile near the position of DDO68-V1 for the epoch
of December 2017.
{\bf Right-hand panel:} Residual signal of the accepted two-Gauss fitting. See parameters of the components in the text.
}
	\label{fig:blendfit_LBV}
 \end{figure}

The results of the two-Gauss fitting in the HST images at the epoch of December 2017 allow one to determine the ratio of
fluxes of the both components, as well as their
relative positions, and to tie their positions to that of the LBV in the HST image on the epoch of May 2010.
Having the instrumental magnitudes in both F606W and F814W filters, derived from the aperture photometry of this
complex object, we use the respective zero-points, as described below, and obtain its Johnson-Cousins
$V$=23.630$\pm$0.005~mag and $I$=23.627$\pm$0.008~mag. From these 'total' magnitudes
we finally derive $V$ and $I$-band magnitudes of the both components, the LBV and its very close neighbour
star ($\delta$X = 2.85~px = 0.114~arcsec, or projected distance of 7~pc).

The respective zero-points are determined from the known $V$ and $I$ magnitudes of the fifty the most luminous
supergiants distributed along the DDO68 'Northern Ring' \HII\ regions. Their magnitudes are measured, as described
in \citet{DDO68LBV}, on the HST images at the epoch of May 2010. We perform the aperture photometry of
a subsample of 35 of these fifty supergiants, which are clearly isolated in the HST images of December 2017. The zero-point
is adopted as a median of all obtained estimates. This statistics is robust, even if some of the used stars
experience small brightness variations between the two HST epochs.

To derive correct absolute magnitudes and colours, we need to adopt the realistic value of extinction.
For further corrections of the stellar absolute magnitudes and colours in Knot~3, we adopt the value of extinction of
$E(B-V)$ = 0.12 and the related $A_{\rm V}$ = 0.36~mag. This value of $E(B-V)$ is substantially larger than
the MW extinction in this direction, $E(B-V)_{\rm MW}$ = 0.016~mag according to  \citet{SF2011}. This larger extinction
corresponds to the average of the three independent estimates of the internal extinction C(H$\beta$) in this
\HII\ region, as determined from the spectra of Knot~3, obtained in the epochs when the LBV was below the
detection limit, as presented in \citet{DDO68, Annibali2019b}.
We suppose that the dust, affecting the relative emission-line fluxes in the \HII\ region, affects as well the total
emission of Knot~3 and the emission of all stars in this region. That is, to derive the absolute magnitudes $M_{\rm V,0}$ with
the account for that extinction, we should use the correction of $\delta~V = -0.36$~mag. The related correction of $E(V-I)$
= 0.165~mag.

For the brighter star, we derive the apparent magnitude $V$ = 23.99$\pm$0.05~mag and colour $(V-I)$ = 0.056$\pm$0.115.
The elevated errors of magnitudes for both components of the close 'pair' (in comparison to the aperture photometry errors)
appear due to the additional uncertainties of the component flux estimate in course of the two-Gauss fitting to the 1D
photometric profile.
The respective absolute $V$-magnitude M$_{\rm V,0}$ = --6.90$\pm$0.05 and colour $(V-I)_{\mathrm 0}$ = --0.109$\pm$0.115.
So that, this star can be assigned to B supergiant.
For LBV (DDO68-V1), we derive the following parameters: $V$ = 25.00$\pm$0.12, $(V-I)$ = --0.146$\pm$0.34.
The respective absolute $V$-magnitude M$_{\rm V,0}$ = --5.89$\pm$0.12 and colour $(V-I)_{\mathrm 0}$ = --0.31$\pm$0.34.
Taken together with the direct estimate of the LBV light ($V$ = 20.05~mag) at the
HST image at the epoch of May 2010 \citep{DDO68LBV}, and its related M$_{\rm V,0}$ = --10.84~mag, this assumes the minimal
total amplitude of DDO68-V1 of $\delta~V \gtrsim$ 5.0~mag.

Since we have for the December 2017 HST data two independent images for each of the filters, we can check possible
light variations of DDO68-V1 on the time scales of 2--3 weeks. For the sum of the two stars in the aperture (BSG+LBV),
we have variations between the individual images of $\delta~V$, $\delta~I$ of 0.01~mag or less, with the formal error of
the individual estimates of 0.0034 mag for $V$-band and 0.0063~mag - for $I$-band. Taking this as an upper limit,
and accounting for the LBV contribution to the total light of this close 'pair' of $\sim$30\%, we estimate the
respective amplitude of variations
of the LBV in this period as $\delta~V \lesssim $0.03~mag.

We should comment, why we adopt for the HST images of 2017 the aperture photometry instead of the common PSF
photometry with the DOLPHOT package. As already mentioned, the HST images in 2010 and 2017 were obtained with the
different instruments, ACS/WFC (Advanced Camera for Surveys) and WFC3/UVIS (Wide Field Camera 3), respectively.
We performed PSF photometry for these two epochs, using the DOLPHOT parameters recommended by \citet{Annibali2019b}.
These authors notice the systematic shifts of 0.03--0.04~mag between the PSF photometry results, corresponding to
ASC and WFC3 instruments. However, our photometry with the same DOLPHOT parameters results in the systematical
shifts up to 0.10--0.15~mag.  We also tried to change DOLPHOT parameters to those recommended in the DOLPHOT manual
and to those from paper by \citet{Wiliams2014}. However, this does not help and the above systematics remains.
Therefore, to remedy the problem, we decide to adopt as the reference values, the photometry with ACS/WFC in the
HST images in 2010. And then, we use the stars with measured magnitudes as the local standards in the images in 2017.

\section{Subtraction of the underlying light of Knot~3 \HII\ region and lightcurve of DDO68-V1}
\label{sec:subtract}

As described above and in the previous works \citep[e.g.][]{DDO68LBV}, we assume, as the first approximation,
that all visible variability of the \HII\ region Knot~3 is due to the (strong) variability of DDO68-V1. To subtract
the non-variable light of the underlying \HII\ region, we need to determine the total magnitude of this region within
the used aperture (here the circle with radius of 2.5~arcsec). As one can see from the lightcurve of region Knot~3 in
Fig.~2 in \citet{DDO68NR}, this level is fainter than the formal minimum of the lightcurve in
$V$-band of $V_{\rm min}$ = 20.20~mag.

To define better the minimal level of the integrated light of Knot~3, in order to use this amount in the analysis
of the ground-based photometry and to derive the residual light of the LBV, we use the following approach.

In Table~\ref{tab:photo_Kn3},  we present all results of the integrated photometry in $B$, $V$ and $R$-bands
for the DDO68 star-forming region Knot~3.
The presented magnitudes include the upward corrections of the measured light which account for the loss of the light
in the fixed-size aperture for different seeings (see Sect.~2.6 in \citet{DDO68NR}).

\begin{figure}
  \centering
\includegraphics[angle=-90,width=8cm, clip=]{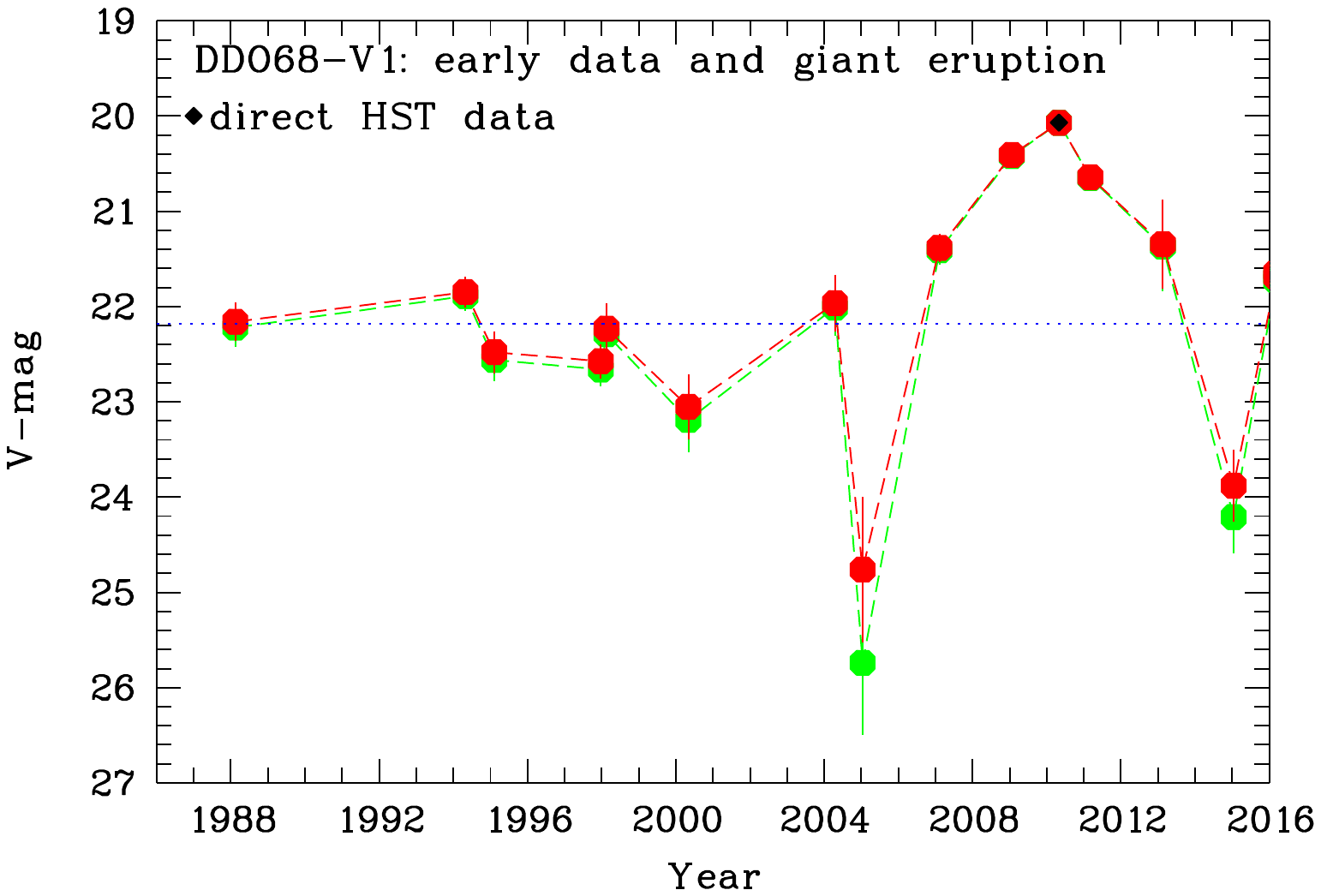}
\includegraphics[angle=-90,width=8cm, clip=]{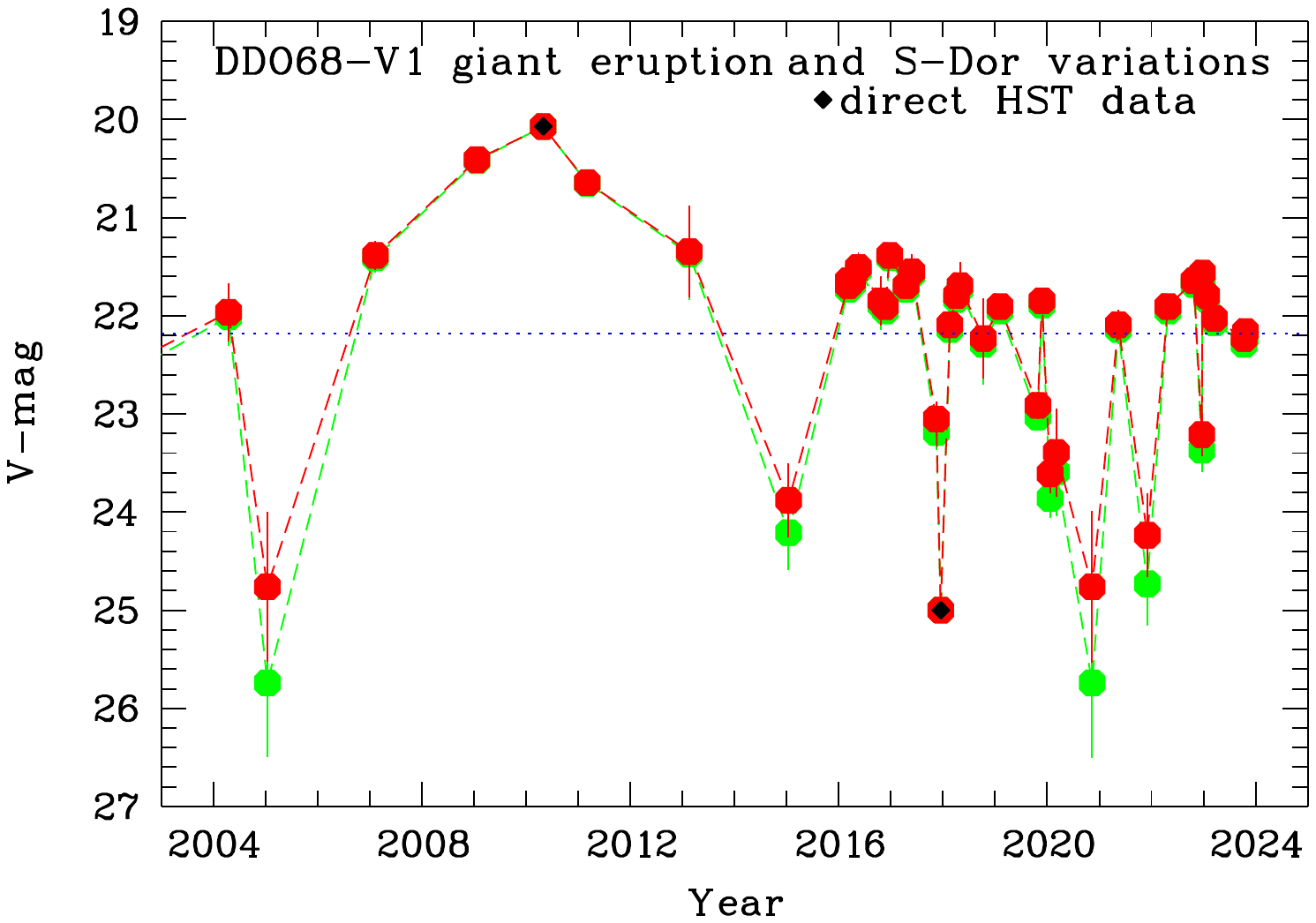}
\includegraphics[angle=-90,width=8cm, clip=]{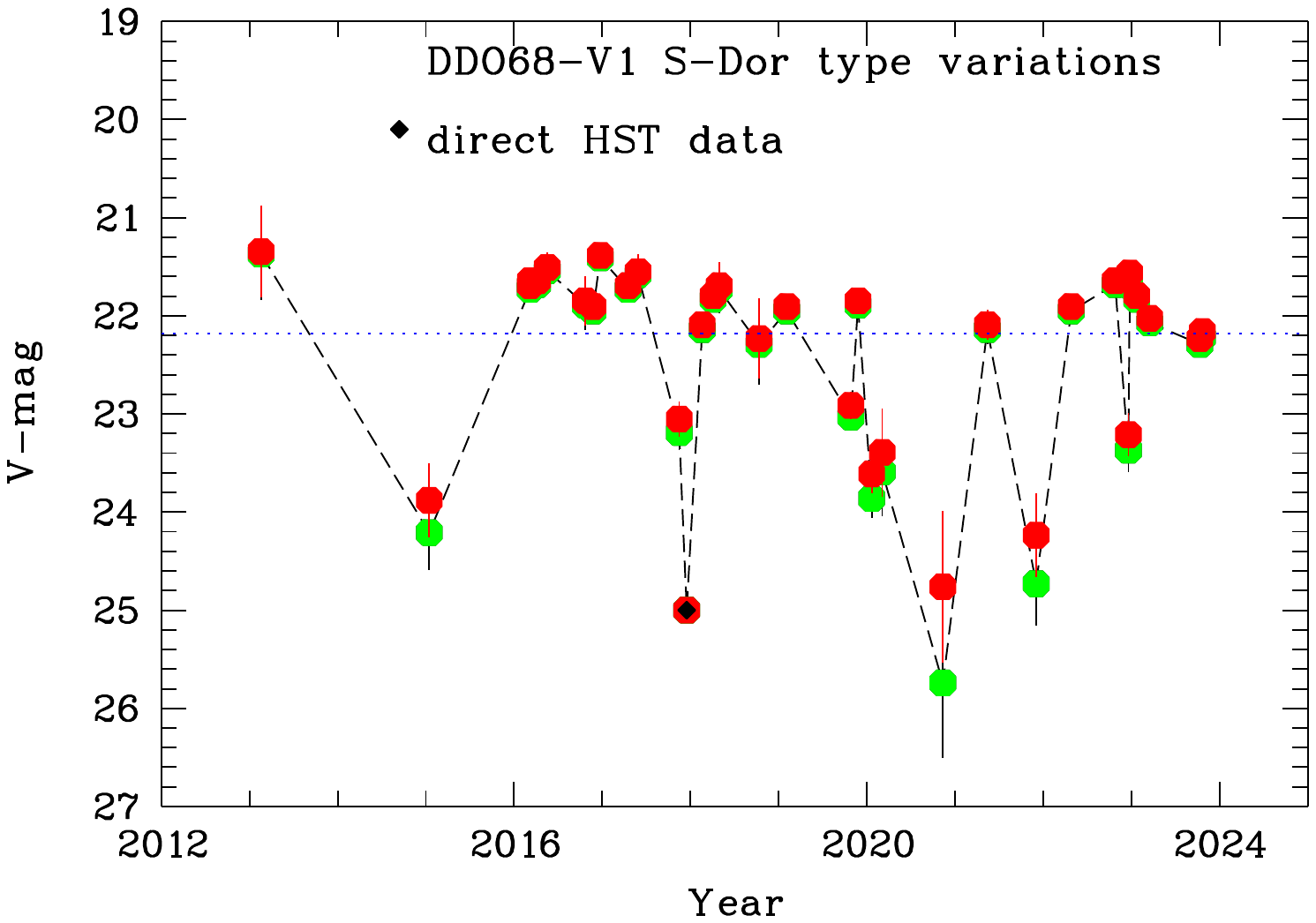}
  \caption{
$V$-band lightcurves  of the LBV DDO68-V1 before, during and after the 'giant eruption' and the follow-up S~Dor
type variations with amplitudes of 3.0 to $\gtrsim$3.5 mag. Green symbols are for the background level of V$_{\rm back}$ =
20.23~mag, while red symbols are for  V$_{\rm back}$ = 20.24~mag.
{\bf Top panel.} The $V$-band lightcurve of the LBV on the archive data and during the giant eruption (1988--2016).
The direct HST measurements are marked by black lozenges.
{\bf Middle panel.} The $V$-band lightcurve for the LBV for giant eruption and afterwards (2004--2024).
{\bf Bottom panel.} The same lightcurve for LBV, for the period of 2014--2024, for a more detailed view of the S~Dor phase.
}
	\label{fig:LBV_eruption}
 \end{figure}

For $V$-band, we have an advantage of the direct measurement in the HST image at epoch of December 2017 of $V$ magnitudes
of both objects in the close optical pair, DDO68-V1 and a blue supergiant neighbour at 0.11~arcsec. Having the total
brightness
of Knot~3 near the bottom of its registrated lightcurve, at the level of V$_{\rm tot}$=20.17--20.19~mag, one can find
the self-consistent level of the background light of the  \HII-region (including the rest stars).
At this level, the faintest observed value of V$_{\rm K3,tot}$ = 20.19$\pm$0.013, after subtraction of the background,
and then of the light of the nearby B supergiant, should give the non-negative signal for LBV.
This criterion gives the estimate of V$_{\rm back}$ = 20.23--20.24~mag.

We adopt for further analysis V$_{\rm back}$ = 20.23. The respective lightcurves in Fig.~\ref{fig:LBV_eruption} are marked
by green symbols, connected by dashed lines. The variant of V$_{\rm back}$ = 20.24 will slightly shift upwards the majority
of DDO68-V1 magnitudes, that will be within their uncertainties. The respective lightcurves are shown by red symbols in the
same figures.  The largest changes for V$_{\rm back}$ = 20.24 appear for the
bottom of the LBV lightcurve, say for the range of V$_{\rm tot}$=20.17--20.19~mag.

The level of V$_{\rm back}$ = 20.23--20.24~mag is further supported by the light of the LBV, derived directly on the HST
images at epoch of December 2017, V$_{\rm LBV}$ = 25.00~mag. This level appears close to several other estimates near
the bottom
of the LBV lightcurve. If we adopt the background level at the fainter brightness, for example, V$_{\rm back}$ = 20.25, all
the  ground-based minima will appear substantially brighter, so that the HST $V$-magnitude of the LBV becomes the only
faintest point in the whole lightcurve. While we can not exclude such a case, this looks less likely than the case
of the HST point and of the other minima to be of the comparable brightness.

A part of the earlier (before year 2016) estimates of Knot~3 brightness is published in \citet{DDO68LBV},
where we confirm the event of the LBV giant eruption.
The photometry and analysis of those images is updated for this paper. Therefore the magnitudes in our
Table~\ref{tab:photo_Kn3} and those in Table~2 from \citet{DDO68LBV} are slightly different. The main factors that lead
to the changes are: 1) a more careful and homogeneously for all dates determined background around Knot~3; 2) the full
account for the loss of the light in the aperture with the fixed radius  due to the variable seeing.

The resulting magnitudes after the background subtraction are presented in Table~\ref{tab:photo_V}. The Cols. 2 and
3 give the $V$-mag for the sum light of the close pair of LBV and BSG, and its error. In Cols. 4 and 5 we give the similar
values for the pure LBV light, after subtraction the light of the BSG from the sum value in Col. 2.

\section{Colour variations of Knot~3 \HII\ region and implication for DDO68-V1}
\label{sec:color.var}

\begin{figure}
  \centering
 \includegraphics[angle=-90,width=8cm, clip=]{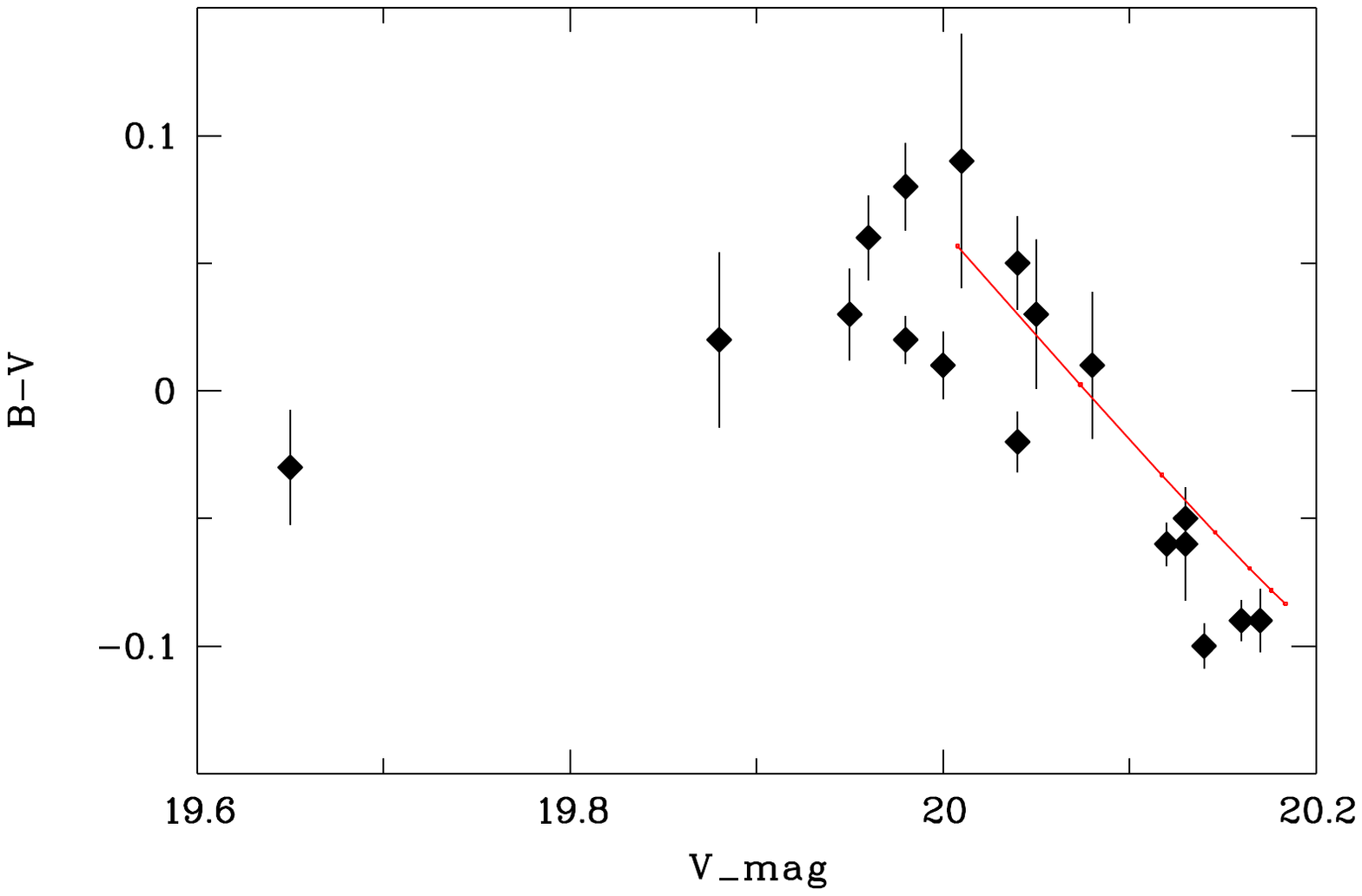}
 \includegraphics[angle=-90,width=8cm, clip=]{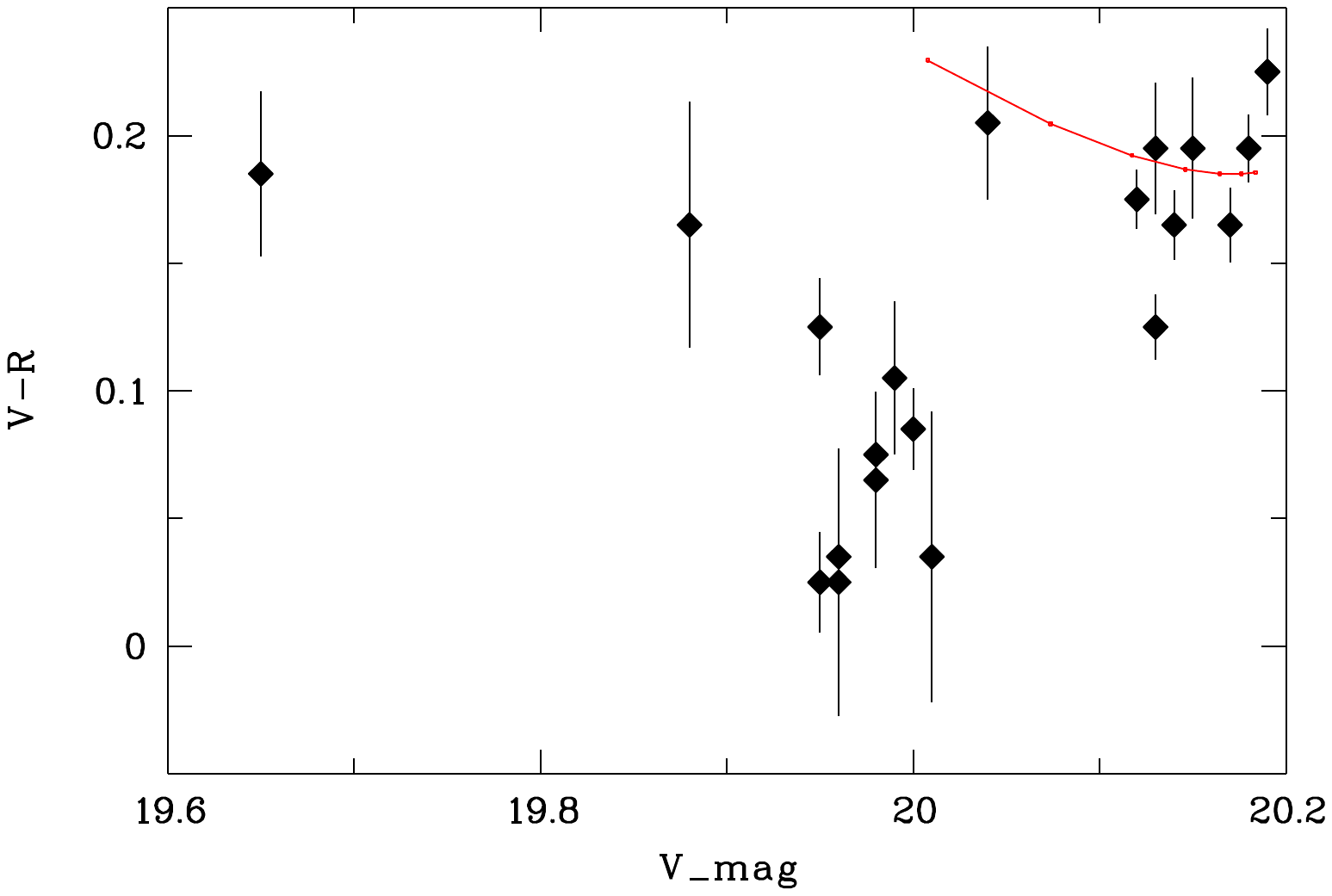}
  \caption{
The colour indices $(B-V)$ (top panel) and $(V-R)$ (bottom panel) of the total Knot~3 light (corrected
   for $E(B-V)$ = 0.12, see text) versus its brightness in $V$ band for all available photometric data.
   The total magnitudes V$_{\rm Kn.3}$ of 20.17--20.19~mag correspond to several deep minima on the lightcurve
   of LBV after the 'giant eruption'. Respectively, the total magnitudes of Knot~3  of
   $V_{\rm Kn.3} \sim$19.96--20.03~mag correspond to the maxima of the LBV lightcurve in this period. The solid
   red lines, overplotted to the both panels, represent the expected changes of Knot~3 colour in the case if
   the $(B-V)$ and $(V-R)$ colours of the LBV change as a linear function of $V$(LBV) between quasi S-Dor
   minimum and maximum. The extremes correspond to an early B (as its $(V-I)$ from the HST data at the epoch
   of December 2017, with $V$ = 25.0~mag) and to late F near the local maxima, when $V$(LBV) $\sim$22.0~mag.
}
	\label{fig:K3_colours}
 \end{figure}

In Fig.~\ref{fig:K3_colours}, we present variations of $(B-V)$ and $(V-R)$ colours of Knot~3, corrected  for
$E(B-V)$ = 0.12, versus its $V$-magnitude.

We assume, that at first approximation, both, the total magnitudes and colour indices of Knot~3, vary only due
to the variable light of the LBV DDO68-V1. Then, the walks of the integrated colours in these plots correspond
to the varying contribution of the LBV light to the total light of Knot~3.

If we exclude the brightest phase, corresponding to the 'giant eruption', the range of $V$(Knot~3)=19.95 to
20.19 corresponds to $V$(LBV) of $\sim$21.7 to $\sim$25.0~mag, that is the range from maxima to minima of
quasi S~Dor variations.
For $(B-V)$, one can see the clear trend of reddening when the LBV
gets brighter and its contribution to the light of the whole Knot~3 (V$_{\rm back}$ = 20.23) reaches 20--25~\%.
This type of colour variations is characteristic of the S~Dor phase. For example, the  $(B-V)$ colour index
changed systematically, getting bluer by $\sim$0.1~mag in the LBV R71 in course of its fading in 1975--1979
by $\sim$1~mag \citep[][and the plot $(B-V$ vs $V$, kindly presented by the referee]{Spoon1994}.

At the same time, $(V-R)$ colour also changes substantially, but gets bluer. While the reddening of LBV along the raising
 luminosity is consistent with the behaviour of Knot~3  $(B-V)$ index,  the variation of $(V-R)$ index looks clearly
inconsistent with the cooling 'atmosphere' of LBV in the course of its expansion.  To illustrate this situation,
we draw by red solid lines the expected changes of the integrated colours of Knot~3 for the simple model, in which
the $(B-V)$ and $(V-R)$ colours of the LBV in this phase vary from late B supergiant to a late F supergiant as the
linear function of its brightness.

Thus, it is evident, that near the local maxima the $(V-R)$ colour of Knot~3 tends to become bluer. One of the
possibilities of such behaviour, could be the diminishing of contribution of the strong emission line H$\alpha$
in the spectrum of DDO68-V1 in $R$-band, what somewhat compensates its reddening in the high state.
The other options include the variation of the H$\alpha$ flux in the related nebulae due to variable ionisation
from the DDO68-V1.

 Due to the related timescales, one can not connect such a behaviour with, for example, variability of emission
H$\alpha$ in the \HII\ region,
since the characteristic time for Hydrogen recombination at N$_{\rm e} \sim$ 10~cm$^{-3}$ is of the order of 10$^4$ years.
In principle, it is possible to examine an option of the compact (fraction of parsec) denser shell as visible around some
Milky Way LBVs (with N$_{\rm e} \sim$ 10$^5$~cm$^{-3}$). If in such a shell, the substantial H$\alpha$ emission makes $(V-R)$
redder near S-Dor minima (aka blue supergiant), then, the switch-off of the ionisation source in the cooler
phases near the maxima, results in the 'fast' recombination and reducing the contribution of H$\alpha$ and its
effect on $(V-R)$ colour.

\clearpage

\begin{table*}
\caption{$B,V,R$ total magnitude estimates for DDO68 SF region Knot~3 adopted from \citet{DDO68NR}
\label{tab:photo_Kn3}
}
\begin{tabular}{llllllll}
\hline
\multicolumn{1}{l}{Date$^{1}$}  &
\multicolumn{1}{c}{$B$} &
\multicolumn{1}{c}{$\sigma_{\rm B}$} &
\multicolumn{1}{c}{$V$} &
\multicolumn{1}{c}{$\sigma_{\rm V}$} &
\multicolumn{1}{c}{$R$} &
\multicolumn{1}{c}{$\sigma_{\rm R}$} &
\multicolumn{1}{l}{Ref. Notes}     \\
 \hline \\[-0.2cm]
19880214   & 20.21 & 0.018 & 20.04 & 0.030 & 19.76 & 0.013 & CA 3.5m                             \\ 
19940502   & --    &  --   & 19.99 & 0.030 & 19.81 & 0.020 & INT 2.54m                           \\ 
19950207   & 20.21 & 0.018 & 20.08 & 0.026 & --    &  --   & NOT 2.56m                           \\ 
19971223   & --    &  --   & 20.09 & 0.010 & --    &  --   & KeckII                              \\ 
19980217   & 20.20 & 0.018 & 20.05 & 0.040 & --    &  --   & KPNO 4m                             \\ 
20000503   & 20.19 & 0.022 & 20.13 & 0.026 & 19.86 & 0.034 & VATT 1.8m                           \\ 
20040416   & 20.22 & 0.050 & 20.01 & 0.057 & 19.90 & 0.072 & SDSS                                \\ 
20050112   & --    &  --   & 20.19 & 0.013 & 19.89 & 0.013 & BTA                                 \\ 
20070209   & --    &  --   & 19.89 & 0.040 & --    &  --   & KPNO 2.1m (2) interpol. from $V-g$  \\ 
20090121   & --    &  --   & 19.55 & 0.008 & --    &  --   & SAO BTA                             \\ %
20100502   & --    &  --   & 19.33 & 0.077 & --    & --    & HST (2)                             \\ 
20110305   & 19.74 & 0.022 & 19.65 & 0.032 & 19.39 & 0.036 & KPNO 0.9m                           \\ 
20130217   & 20.02 & 0.034 & 19.88 & 0.048 & 19.64 & 0.044 & KPNO 0.9m                           \\ 
20150114   & 20.20 & 0.012 & 20.17 & 0.015 & 19.93 & 0.011 & SAO BTA                             \\
20160307   & --    & --    & 19.96 & 0.013 & --    & --    & SAO 1m                              \\ 
20160308   & 20.10 & 0.018 & 19.95 & 0.019 & 19.75 & 0.023 & SAO 1m                              \\ 
20160407   & --    & --    & 19.95 & 0.015 & --    & --    & SAO 1m                              \\ 
20160517   & --    & --    & 19.92 & 0.045 & --    & --    & SAO 1m                              \\ 
20161022   & --    & --    & 19.99 & 0.024 & --    & --    & SAO 1m                              \\ 
20161124   & 20.13 & 0.013 & 20.00 & 0.016 & 19.84 & 0.024 & SAO 1m                              \\ 
20161224   & --    & --    & 19.89 & 0.019 & --    & --    & SAO 1m                              \\ 
20161231   & --    & --    & 19.99 & 0.011 & --    & --    & SAO 1m                              \\ 
20170418   & --    & --    & 19.96 & 0.008 & --    & --    & CMO 2.5m                            \\ 
20170529   & --    & --    & 19.93 & 0.044 & --    & --    & SAO 1m                              \\ 
20171116   & 20.20 & 0.011 & 20.13 & 0.013 & 19.93 & 0.011 & SAO BTA                             \\ 
20180219   & --    & --    & 20.03 & 0.009 & --    & --    & CMO 2.5m                            \\ 
20180405   & 20.18 & 0.017 & 19.98 & 0.035 & 19.84 & 0.022 & SAO 1m                              \\ 
20180430   & --    & --    & 19.96 & 0.052 & 19.86 & 0.050 & SAO 1m                              \\ 
20181011   & --    & --    & 20.05 & 0.066 & --    & --    & SAO 1m                              \\ 
20190118$^3$& --   & --    & (20.01)& 0.043& 19.77 & 0.018 & CMO 2.5m                            \\ 
20190203   & --    & --    & 20.00 & 0.018 & --    & --    & SAO 1m                              \\ 
20191026   & 20.18 & 0.009 & 20.12 & 0.012 & 19.87 & 0.011 & SAO BTA                             \\ 
20191125   & --    & --    & 19.99 & 0.021 & --    & --    & SAO 1m                              \\ 
20200119   & --    & --    & --    & --    & 19.87 & 0.008 & SAO BTA                             \\ 
20200120   & 20.19 & 0.008 & 20.16 & 0.009 & --    & --    & SAO BTA                             \\ 
20200304   & --    & --    & 20.15 & 0.028 & 19.88 & 0.030 & SAO 1m                              \\ 
20200426   & --    & --    & --    & --    & 19.89 & 0.011 & SAO BTA                             \\ 
20201111   & --    & --    & 20.19 & 0.017 & 19.89 & 0.011 & SAO BTA                             \\ 
20210514   & --    & --    & 20.03 & 0.018 & --    & --    & SAO 1m                              \\ 
20210516   & --    & --    & --    & --    & 19.93 & 0.027 & SAO 1m                              \\ 
20211202   & --    & --    & 20.18 & 0.013 & 19.91 & 0.013 & SAO BTA                             \\ 
20220426   & --    & --    & 20.00 & 0.021 & --    & --    & SAO 1m                              \\ 
20221025   & --    & --    & 19.95 & 0.020 & 19.85 & 0.026 & SAO 1m                              \\ 
20221220   & 20.16 & 0.009 & 20.14 & 0.014 & 19.90 & 0.010 & SAO BTA                             \\ 
20221225   & 20.14 & 0.017 & 19.96 & 0.013 & 19.85 & 0.015 & SAO 1m                              \\ 
20230123   & 20.12 & 0.009 & 19.98 & 0.011 & 19.83 & 0.012 & SAO 1m                              \\ 
20230318   & --    & --    & 20.02 & 0.014 & --    & --    & SAO 1m                              \\ 
20231012   & --    & --    & 20.05 & 0.017 & --    & --    & SAO 1m                              \\ 
20231022   & 20.14 & 0.012 & 20.04 & 0.018 & --    & --    & SAO BTA                             \\ 
\\[-0.25cm] \hline \\[-0.2cm]
 \multicolumn{8}{l}{(1) Date in the format YYYYMMDD. (2) adopted from \citet{DDO68LBV}. } \\
 \multicolumn{8}{l}{(3) V-band mag is adopted from R-band with the average V--R for similar R-mag.} \\
\end{tabular}
\end{table*}

\begin{table*}
\centering
\caption{$V$-band magnitudes for the sum of LBV and the close B-supergiant (denoted as B+LBV), and for the pure\\
LBV, after subtracting the B-supergiant light ($V$=23.99) from the sum. Adopted background: $V_{\rm back}$=20.23~mag.
\label{tab:photo_V}
}
\begin{tabular}{lllll}
\hline
\multicolumn{1}{l}{Date}  &
\multicolumn{1}{c}{$V$(B+LBV)} &
\multicolumn{1}{c}{$\sigma_{\rm V}$(B+LBV)} &
\multicolumn{1}{c}{$V$(LBV)} &
\multicolumn{1}{c}{$\sigma_{\rm V}$(LBV)} \\
 \hline \\[-0.2cm]
19880214  & 22.026& +0.20,-0.18 & 22.220& +0.23,-0.20  \\  
19940502  & 21.747& +0.16,-0.14 & 21.894& +0.19,-0.16  \\  
19950207  & 22.303& +0.22,-0.18 & 22.561& +0.26,-0.22  \\  
19971223  & 22.383& +0.14,-0.13 & 23.664& +0.22,-0.18  \\  
19980217  & 22.090& +0.29,-0.24 & 22.297& +0.34,-0.27  \\  
20000503  & 22.769& +0.32,-0.26 & 23.195& +0.47,-0.34  \\  
20040416  & 21.851& +0.35,-0.23 & 22.014& +0.40,-0.30  \\  
20050112  & 23.794& +0.40,-0.38 & 25.745& +1.34,-1.34  \\  
20070209  & 21.316& +0.16,-0.08 & 21.413& +0.17,-0.15  \\  
20090121  & 20.380& +0.02,-0.02 & 20.420& +0.02,-0.02  \\  
20100502  & 20.050& +0.08,-0.08 & 20.077& +0.08,-0.08  \\  
20110305  & 20.608& +0.08,-0.08 & 20.657& +0.08,-0.08  \\  
20130217  & 21.280& +0.67,-0.45 & 21.373& +0.88,-0.53  \\  
20150114  & 23.344& +0.33,-0.24 & 24.213& +2.50,-0.73  \\  
20160307  & 21.603& +0.11,-0.10 & 21.731& +0.13,-0.12  \\  
20160308  & 21.558& +0.17,-0.15 & 21.681& +0.21,-0.20  \\  
20160407  & 21.558& +0.11,-0.10 & 21.681& +0.13,-0.12  \\  
20160517  & 21.423& +0.32,-0.26 & 21.540& +0.38,-0.30  \\  
20161022  & 21.747& +0.26,-0.22 & 21.894& +0.33,-0.27  \\  
20161124  & 21.798& +0.13,-0.12 & 21.953& +0.17,-0.15  \\  
20161224  & 21.316& +0.13,-0.12 & 21.413& +0.15,-0.14  \\  
20170418  & 21.603& +0.05,-0.05 & 21.731& +0.06,-0.06  \\  
20170529  & 21.473& +0.29,-0.24 & 21.586& +0.35,-0.27  \\  
20171116  & 22.769& +0.77,-0.46 & 23.195& +1.01,-1.14  \\  
20171216  & 23.630& +0.01,-0.01 & 25.000& +0.12,-0.12  \\  
20180219  & 21.965& +0.07,-0.06 & 22.148& +0.09,-0.08  \\  
20180405  & 21.697& +0.31,-0.25 & 21.837& +0.39,-0.30  \\  
20180430  & 21.603& +0.43,-0.32 & 21.731& +0.55,-0.38  \\  
20181011  & 22.090& +0.54,-0.38 & 22.297& +0.82,-0.48  \\  
20190203  & 21.798& +0.12,-0.11 & 21.953& +0.15,-0.13  \\  
20191026  & 22.660& +0.15,-0.13 & 23.038& +0.35,-0.26  \\  
20191125  & 21.747& +0.12,-0.11 & 21.894& +0.15,-0.13  \\  
20200120  & 23.171& +0.45,-0.32 & 23.861& +0.70,-0.70  \\  
20200304  & 23.021& +1.77,-0.66 & 23.593& +0.70,-0.70  \\  
20201111  & 23.794& +1.77,-1.20 & 25.745& +2.70,-0.70  \\  
20210514  & 21.965& +0.16,-0.14 & 22.148& +0.21,-0.18  \\  
20211202  & 23.547& +1.77,-1.20 & 24.732& +0.70,-0.70  \\  
20220426  & 21.798& +0.12,-0.11 & 21.953& +0.15,-0.13  \\  
20221025  & 21.558& +0.45,-0.34 & 21.681& +0.76,-0.48  \\  
20221220  & 22.888& +0.23,-0.19 & 23.377& +0.50,-0.34  \\  
20221225  & 21.603& +0.06,-0.06 & 21.692& +0.07,-0.07  \\  
20230123  & 21.697& +0.34,-0.27 & 21.837& +0.44,-0.29  \\  
20230318  & 21.907& +0.19,-0.17 & 22.079& +0.24,-0.20  \\  
20231012  & 22.090& +0.14,-0.13 & 22.297& +0.18,-0.16  \\  
20231022  & 22.026& +0.10,-0.10 & 22.220& +0.13,-0.12  \\  
 \hline \\[-0.2cm]
\end{tabular}
\end{table*}

\end{appendix}

\begin{thebibliography}{99}

\bibitem[\protect\citeauthoryear{Abazajian et al.}{2009}]{DR7}
       Abazajian K.N., et al., 
       2009, ApJS, 182, 543

\bibitem[\protect\citeauthoryear{Aghakhanloo et al.}{2023}]{Aghak2023a}
	Aghakhanloo M., Smith N., Milne P., et al.,
	2023, MNRAS, 521, 1941


\bibitem[\protect\citeauthoryear{Allan et al.}{2020}]{Allan2020}
   Allan A.P., Groh J.H., Mehner A., Smith N., Boian I., Farrel E.J., Andrews J.E.
   2020, MNRAS, 496, 1902


\bibitem[\protect\citeauthoryear{Annibali et al.}{2019a}]{Annibali2019a}
	Annibali F., Bellazzini M., Correnti M., Sacchi E., Tosi M.,
	Cignioni M., Aloisi A., et al.,
	2019a, ApJ, 883:19 (12pp)

\bibitem[\protect\citeauthoryear{Annibali et al.}{2019b}]{Annibali2019b}
	Annibali F., La Torre V., Tosi M., et al.,
	2019b, MNRAS, 482, 3892



\bibitem[\protect\citeauthoryear{Bomans \& Weis}{2011}]{Bomans11}
  Bomans D.J., and Weis K.,
  2011, Bulletin de la Societe Royale des Sciences de Li\'ege, 80, 341



\bibitem[\protect\citeauthoryear{Chen et al.}{2015}]{PARSEC}
     Chen Y., Bressan A., Girardi L., Marigo P., Kong X., Lanza A.,
	2015, MNRAS, 452, 1068


\bibitem[\protect\citeauthoryear{Crowther}{2004}]{Crowther04}
     Crowther P., 2004, in Evolution of Massive Stars, Mass Loss and Winds,
   ed. M.Heydary-Malayeri, Ph.Stee \& J.-P. Zahn, EAS Publications Series, v.13, 119


\bibitem[\protect\citeauthoryear{Drissen et al.}{2001}]{Drissen01}
  Drissen L., Crowther P.A., Smith L.J., Robert C., Roy J.-R., Hillier D.J.,
  2001, ApJ, 546, 481

\bibitem[\protect\citeauthoryear{Ducati et al.}{2001}]{Ducati01}
  Ducati J.R., Bevilacqua C.M., Rembold S.B., Ribeiro D.,
  2001, ApJ, 558, 309

\bibitem[\protect\citeauthoryear{Ekta, Chengalur, Pustilnik}{Ekta et al.}
	{2008}]{Ekta08}
	Ekta, Chengalur J.N., Pustilnik S.A., 2008, \mnras, 391, 881

\bibitem[\protect\citeauthoryear{Eldridge, Stanway}{2022}]{Eldridge22}
  Eldridge J.J., Stanway E.E., 2022, ARAA, 60, 455



\bibitem[\protect\citeauthoryear{Garcia et al.}{2021}]{Garcia21}
  Garcia M., Evans C.J., Bestenlehner J.M. et al., 2021, Experimental Astronomy, 51:887-911


\bibitem[\protect\citeauthoryear{Grassitelli et al.}{2021}]{Grassitelli2021}
  Grassitelli L., Langer N., Mackey J. Gr\"afener G., Grin N.J.,
  Sander A.A.C., Vink J.S., 2021, A\&A, 647, A99

\bibitem[\protect\citeauthoryear{Gull et al.}{2022}]{Gull22}
  Gull M., Weisz D.R., Senchyna P., et al., 2022, ApJ, 941, 206

\bibitem[\protect\citeauthoryear{Guseva et al.}{2022}]{Guseva2022}
  Guseva N.G., Thuan T.X., Izotov Y.I., 2022, MNRAS, 512, 4298


\bibitem[\protect\citeauthoryear{Humphreys \& Davidson}{1994}]{HD94}
       Humphreys R.M., \& Davidson K., 1994, PASP, 106, 1025

\bibitem[\protect\citeauthoryear{Izotov \& Thuan}{2007}]{IT07}
	Izotov Y.I., Thuan T.X., 2007,  \apj,  665, 1115

\bibitem[\protect\citeauthoryear{Izotov \& Thuan}{2009}]{IT09}
	Izotov Y.I., Thuan T.X., 2009,  \apj, 690, 1797




\bibitem[\protect\citeauthoryear{Kniazev, Gvaramadze, Berdnikov}{Kniazev et al.}{2016}]{Kniazev2016}
    Kniazev A.Y., Gvaramadze V.V., Berdnikov L.N.,
    2016, MNRAS, 459, 3068


\bibitem[\protect\citeauthoryear{Lorenzo et al.}{2022}]{Lorenzo22}
   Lorenzo M., Garcia M., Najarro F., Herrero A., Cervi\~no M., Castro N.,
   2022, MNRAS, 516, 4164

\bibitem[\protect\citeauthoryear{Lupton et al.}{2005}]{Lupton05}
 \mbox{Lupton~R.,~et~al.~2005}, https://www.sdss3.org/dr8/algorithms/ \\sdssUBVRITransform.php\#Lupton2005

\bibitem[\protect\citeauthoryear{Mahy et al.}{2022}]{Mahy2022}
 Mahy L., Lanthermann C., Hutsem\'{e}kers D., et al.,
    2022, A\&A, 657, id.A4, 15pp

\bibitem[\protect\citeauthoryear{Makarov et al.}{2017}]{Makarov17}
    Makarov D.I., Makarova L.N., Pustilnik S.A., Borisov S.B.,
    2017, MNRAS, 466, 556



\bibitem[\protect\citeauthoryear{Pastorello et al}{2010}]{Pastorello2010}
 Pastorello A., Boticella M.T., Trundle C., et al., 2010, MNRAS, 408, 181


\bibitem[\protect\citeauthoryear{Petit, Drissen \& Crowther}{Petit et al.}{2006}]{Petit06}
   Petit V.,  Drissen L.,  Crowther P.A., 2006, AJ, 132, 1756

\bibitem[\protect\citeauthoryear{Petrov, Vink \& Gr\"afener}{2016}]{Petrov2016}
      Petrov B., Vink Y., Gr\"afener G., 2016, MNRAS, 458, 1999-2011

\bibitem[\protect\citeauthoryear{Polcaro et al.}{2016}]{Polcaro2016}
      Polcaro V.F., Maryeva O., Nesci R., et al., 2016, AJ, 151, 149

\bibitem[\protect\citeauthoryear{Pustilnik, Tepliakova}{2011}]{PaperI}
      Pustilnik S.A., Tepliakova A.L., 2011, MNRAS, 415, 1188

\bibitem[\protect\citeauthoryear{Pustilnik, Kniazev, \& Pramskij}{Pustilnik et al}{2005}]
  {DDO68} Pustilnik S., Kniazev A., Pramskij A., 2005, \aap,
  443, 91

\bibitem[\protect\citeauthoryear{Pustilnik et al.}{2008}]{LBV}
     Pustilnik S.A., Tepliakova A.L., Kniazev A.Y., Burenkov A.N.,
     2008, MNRAS Lett., 388, L24

\bibitem[\protect\citeauthoryear{Pustilnik et al.}{2017}]{DDO68LBV}
   Pustilnik S.A., Makarova L.N., Perepelitsyna Y.A., Moiseev A.V.,
   Makarov D.I., 2017, MNRAS, 465, 4985

\bibitem[\protect\citeauthoryear{Pustilnik et al.}{2024}]{DDO68NR}
  Pustilnik S.A., Perepelitsyna Y.A., Vinokurov A.S., et al.
  2024, Astrophys.Bull., 79, 594 (arXiv:2411.07393)

\bibitem[\protect\citeauthoryear{Sacchi et al.}{2016}]{Sacchi16}
 Sacchi E., Annibali F., Cignoni M., et al.
  2016, ApJ, 830, 3

\bibitem[\protect\citeauthoryear{Sanyal et al.}{2017}]{Sanyal2017}
 Sanyal D., Langer N., Szecsi D., Yoon S.-C., Grassitelli L.,
 2017, A\&A, 597, A71

\bibitem[\protect\citeauthoryear{Schlafly, Finkbeiner}{2011}]{SF2011}
  Schlafly E.F., \&  Finkbeiner D.P., 2011, ApJ, 737, article id. 103



\bibitem[\protect\citeauthoryear{Smith}{2014}]{Smith2014}
	Smith N., 2014, ARAA, 52, 487

\bibitem[\protect\citeauthoryear{Smith}{2017}]{Smith2017}
	Smith N., 2017, Phil. Trans. R. Soc. A, 375, 20160268

\bibitem[\protect\citeauthoryear{Smith \& Owocki}{2006}]{Smith2006}
	Smith N., \& Owocki S.P., 2006, ApJ, 645, L45

\bibitem[\protect\citeauthoryear{Smith et al.}{2016}]{Smith2016}
   Smith N., Andrews J.E., Mauerhan J.C., Zheng W.K., Filippenko A.V.,
   Graham M.L., Milne P.,   2016, MNRAS, 455, 3546

\bibitem[\protect\citeauthoryear{Smith et al.}{2019}]{Smith2019}
   Smith N., Aghakhanloo M., Murphy J.W., Drout M.R., Stassun K.G., Groh J.H.,
   2019, MNRAS, 488, 1760

\bibitem[\protect\citeauthoryear{Smith et al.}{2020}]{Smith2020}
	Smith N., Andrews J., Moe M., et al., 2020, MNRAS, 492, 5897

\bibitem[\protect\citeauthoryear{Spoon et al.}{1994}]{Spoon1994}
  Spoon H.W.W., de Koter A., Sterken C., Lamers H.J., Stahl O., 1994, A\&AS, 106, 141

\bibitem[\protect\citeauthoryear{Sterken}{2003}]{Sterken2003}
	Sterken C., 2003, in ASP Conference Series, 292, 437

\bibitem[\protect\citeauthoryear{Szecsi et al}{2015}]{Szecsi15}
      Szecsi D., Langer N., Yoon S.-C., Debashis S., de Mink S., Evans C.J.,
     Dermine T., 2015, A\&A, 581, A15




\bibitem[\protect\citeauthoryear{Vink}{2022}]{Vink2022}
  Vink J.S., 2022, ARAA, 60, 203

\bibitem[\protect\citeauthoryear{Vink et al.}{2023}]{Vink_ULLYSSES}
  Vink J., Mehner A., Crowther P., et al., 2023, A\&A, 675, A154

\bibitem[\protect\citeauthoryear{Walborn et al.}{2008}]{Walborn2008}
   Walborn N.,  Stahl O., Carmen R., et al., 2008,  ApJ, 683, L33

\bibitem[\protect\citeauthoryear{Weis, Bomans}{2020}]{Weis2020}
  Weis K., Bomans D.J., 2020, Galaxies, 8, 20

\bibitem[\protect\citeauthoryear{Wegner}{1994}]{Wegner94}
  Wegner W., 1994,  MNRAS, 270, 279

\bibitem[\protect\citeauthoryear{Wiliams}{2014}]{Wiliams2014}
   Williams B.F., Lang D., Dalcanton J.J., et al. 2014, ApJS, 215, 9

\end{thebibliography}
\end{document}